\documentclass[%
pre, %
twocolumn, %
showpacs, %
amssymb, amsmath, %
nobibnotes, %
aps]{revtex4}
\usepackage{bm}%

\usepackage{epsfig}
\usepackage{latexsym}
\usepackage{amsthm}
\usepackage{amsfonts}
\usepackage{amsbsy}
\usepackage{here}
\usepackage{psfrag}

\expandafter\ifx\csname package@font\endcsname\relax\else
 \expandafter\expandafter
 \expandafter\usepackage
 \expandafter\expandafter
 \expandafter{\csname package@font\endcsname}%
\fi

\begin{document}

\title{Bubble generation in a twisted and bent DNA-like model}%

\author{P.V. Larsen}%
\email{P.V.Larsen@mat.dtu.dk}
\affiliation{Informatics and Mathematical Modelling and Department of Mathematics,
	     Technical University of Denmark, 
	     DK-2800 Kgs. Lyngby, Denmark}

\author{P.L. Christiansen}%
\affiliation{Informatics and Mathematical Modelling and Department of Physics,
             Technical University of Denmark, DK-2800 Kgs. Lyngby, Denmark}

\author{O. Bang}%
\affiliation{Research Center COM and Informatics and Mathematical Modelling,
		Technical University of Denmark, DK-2800 Kgs. Lyngby, 
		Denmark}

\author{J.F.R. Archilla}
\affiliation{Departamento de Fisica Aplicada I, Universidad de Sevilla, 
	     Avda. Reina Mercedes s/n, 41012 Sevilla, Spain}

\author{Yu.B. Gaididei}%
\affiliation{Bogolyubov Institute for Theoretical Physics, 03143 Kiev, 
Ukraine}


\begin{abstract}
The DNA molecule is modeled by a parabola embedded chain with 
long-range interactions between twisted base pair dipoles. A  
mechanism for bubble generation is presented and investigated in two 
different configurations. 
Using random normally distributed initial 
conditions to simulate thermal fluctuations, a relationship 
between bubble generation, twist and curvature is established.
An analytical approach supports the numerical results.
\end{abstract}
\date{\today}%
\maketitle
%
\section{\label{sec:intro}Introduction}
In recent years, biomolecular modeling has received an ever increasing amount 
of attention, especially focused on the DNA molecule as well as protein 
structures. The basic structure of DNA is fairly well understood since the 
discovery of Crick and Watson \cite{Nat_171}, but 
it is becoming increasingly apparent that structure alone does not explain its 
complex functionality sufficiently \cite{Und DNA, Saenger, Yaku, Reiss}.\par
An example is the mechanism leading to \emph{bubble generation} in DNA, in 
which the two polypeptide strands open to allow replication of the molecule, 
processing of proteins or complete strand separation (denaturation). 
Thermal fluctuations at physiological temperatures and nonlinear localizations 
are expected to produce
bubbles when geometrical features, such as twist and curvature, are taken 
into account.\par
In initial works investigating the denaturation bubble, the geometrical 
features of the molecule were essentially neglected and energy localization 
was mostly attributed to inhomogeneities and impurities in the lattice chain, 
which may model the action of an enzyme \cite{PRA 40, PRE 53/1, PRE 67, 
PRE 55/4, JPA 35_10519, Muto, PRE 49}, or nonlinear excitations \cite{PLA 154, 
Bang/Peyrard, PRE 55/4, JBP 25, PRL 70, PRE 51}. Also, discreteness plays an 
important role for the localization of these excitations. 
The inhomogeneities have been modeled by different masses at various chain 
sites \cite{PRE 49, PRE 53/1, PRE 67, PRA 42}, by conformational defects 
\cite{PRE 51} or by changes in the coupling between molecular sites 
\cite{PRE 55/4, PRE 67}. Also, different on-site potentials 
\cite{JPA 35_10519, PRA 42, PRA 44, PNAS 100} have been used as 
inhomogeneities, corresponding to the 
different number of hydrogen bonds between the two strands. In DNA, the AT
base pair connects through two hydrogen bonds, whereas the CG base pair has three.
It has to some extent been experimentally verified that bubbles form at 
AT-rich sections of the DNA molecule \cite{PRL 90_138101}, but recent work 
\cite{NAR 32} also suggests that other mechanisms are involved.\par
Recently, both long-range dipole-dipole interaction \cite{PD 163, PLA 249}, 
helicity \cite{PLA 253, JBP 24} and curvature 
\cite{EPL 59, JPA 34_8465, Cond 13} have been included in the nonlinear 
transport theory, as well as combinations of these effects 
\cite{JPA 35_8885, PRE 66, JPA 34_6363, PRE 62, PLA 299}. It has been shown 
that chain geometry induces effects similar to those of impurities 
\cite{EPL 59, JPA 34_8465, JPA 35_8885, Cond 13}.\par
In biological environments, thermal fluctuations are always present and have 
been considered in Refs.~\cite{PD 119, PRE 60, Muto, PRE 47/1, PD 57, 
PRE 63_021901}, for example. In these references it was shown that solitons or 
discrete breathers can be generated from or exist among random thermal 
fluctuations.\par
The aim of the present work is to study an augmented Peyrard-Bishop model of 
the DNA molecule \cite{PRL 62}. We include both dipole-dipole long-range 
interaction and chain geometry in the form of a rigid, parabola embedded 
chain. Elaborating on earlier work \cite{Funnel}, we show how both chain 
curvature and twist can initiate bubble generation in the DNA molecule. 
We consider two different chain configurations and 
randomly 
distributed initial conditions, modeling physiological temperatures.\par
In Sec.~\ref{sec:model}, we introduce the model Hamiltonian and equations of 
motion and discuss relevant parameter values as well as the dipole interaction 
and chain geometry. In Sec.~\ref{sec:num}, numerical investigations are 
performed and the results are supported by an analytical approach in 
Sec.~\ref{sec:anal}. Finally, Sec.~\ref{sec:concl} contains a summary and a
discussion.
\section{\label{sec:model}Model}
\begin{figure}
 \centerline{
   \includegraphics[width=40mm,angle=0]{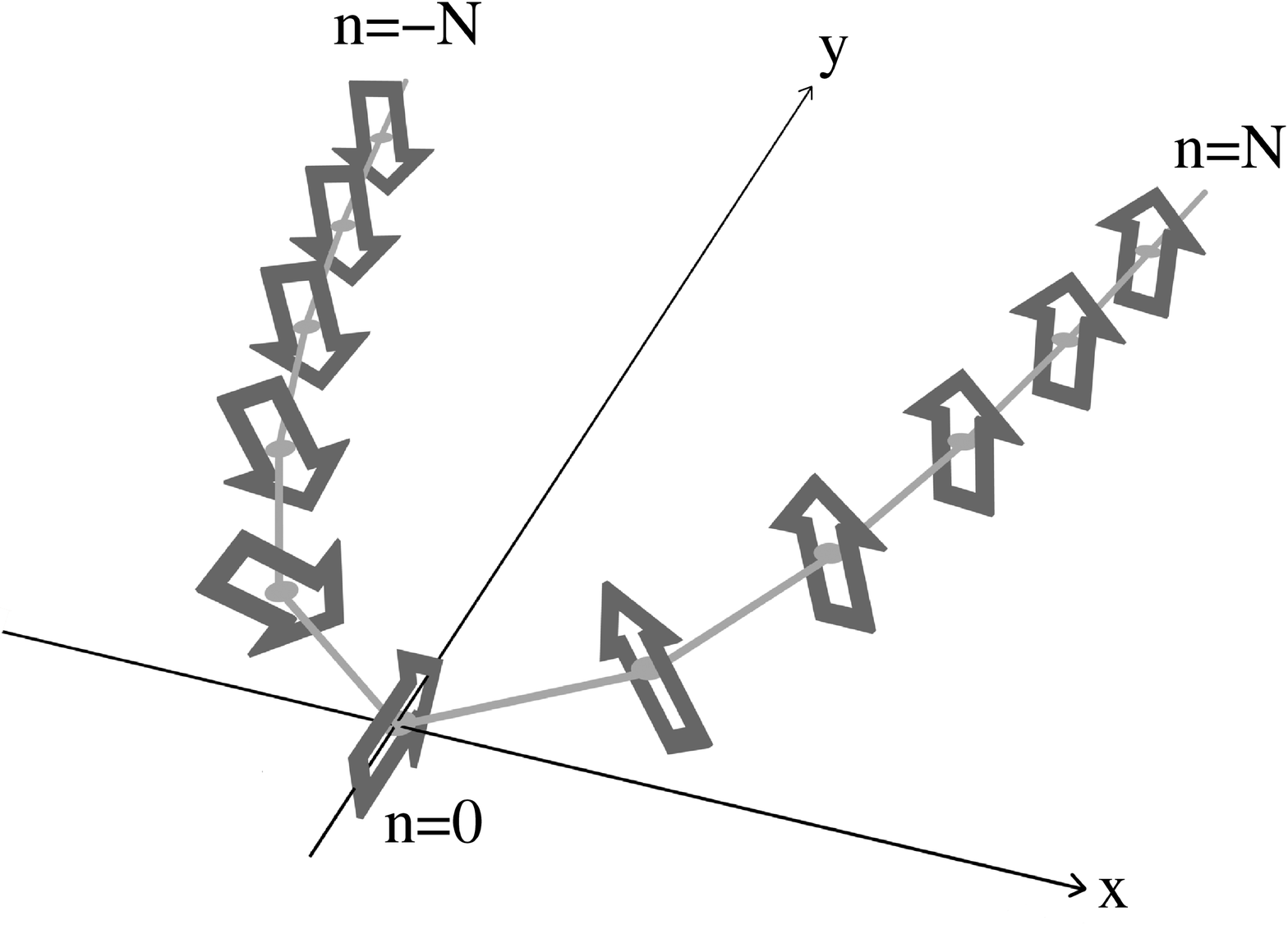}
   \hspace{0mm} 
   \includegraphics[width=40mm,angle=0]{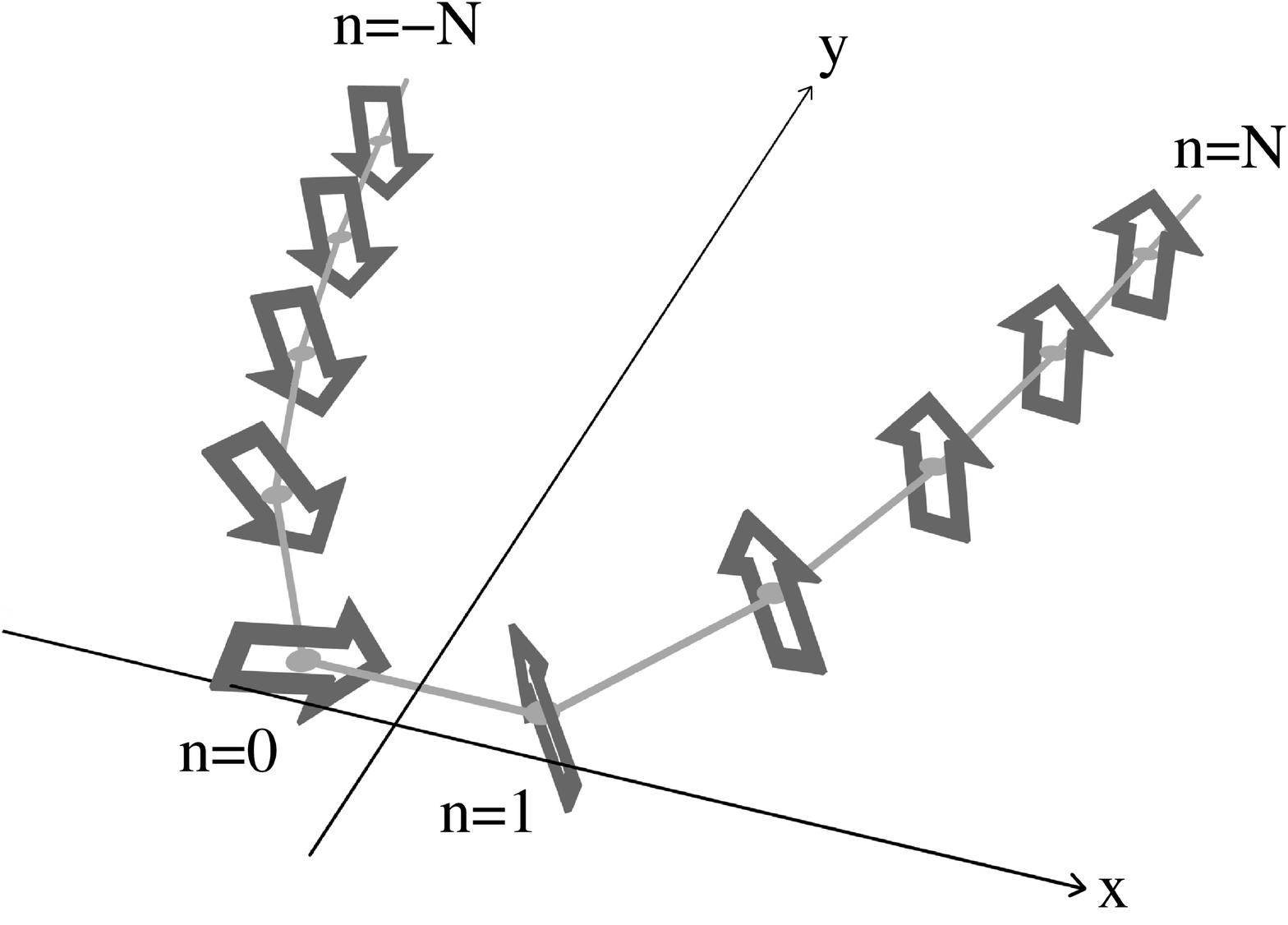}
 }
 \caption{\label{fig:dnachain} DNA chain embedded on a parabola in the 
 $xy$ plane. Sites $(x_n, y_n)$ with $y_n=(\kappa/2) (x_n-\gamma)^2$, indicated by 
 light grey dots. Base pair dipoles, orthogonal to the parabola, shown as 
 dark grey arrows. Curvature $\kappa = 2$ and twist $\tau=1$. 
 The $z$ axis (not shown) forms a right-handed system 
 with $x$ and $y$ axes. Left, the on-site case, $\gamma=0$. 
 Right, the intersite case, $\gamma=1/2$.}
\end{figure}

\begin{figure}[h] 
 \scriptsize
  \psfrag{x1}[cc][cc]{$-2$}
  \psfrag{x2}[cc][cc]{$0$}
  \psfrag{x3}[cc][cc]{$2$}
  \psfrag{x4}[cc][cc]{$4$}
  \psfrag{x5}[cc][cc]{$6$}
  \psfrag{y1}[Br][br]{$0$}
  \psfrag{y2}[Br][br]{$2$}
  \psfrag{y3}[Br][br]{$4$}
  \psfrag{un}[tt][tc]{Displacement, $u_n$}
  \psfrag{Morse}[Bc][Bc]{$V_{\textrm{Morse}}(u_n)$}
 \normalsize
  \centerline{
    \epsfig{file=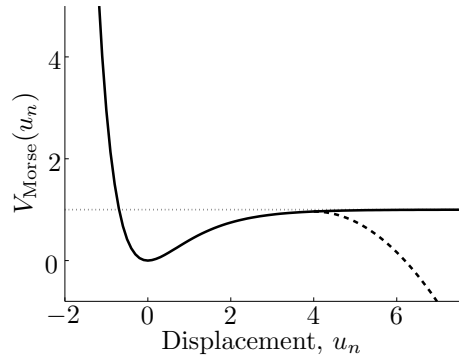, width=6 cm, angle=0} } 
  \caption{The Morse potential 
  	$V_{\textrm{Morse}}(u_n) = \left( e^{-u_n} -1 \right)^2$ 
	  (solid curve). Effective potential, Eq.~(\ref{eq:veff}), in the 
  analytical approximation (dashed curve).}
  \label{fig:Morse}
\end{figure}

We consider parabola embedded chains, as 
illustrated in Fig.~\ref{fig:dnachain}. The base pair sites are 
embedded along a parabola in the $xy$ plane at uniform distances. The dipole 
moments of the base pairs, represented by arrows, are orthogonal to the 
parabola and twisted in the orthogonal plane. Within the $n$th 
base pair, the deviation from the equilibrium transverse distance between the 
bases is denoted $u_n$ \cite{PRL 62}.
The intrasite dynamics of the base pairs is governed by a Morse potential, see 
Fig.~\ref{fig:Morse}, and we assume a harmonic intersite coupling between 
neighboring base pairs. The coupling---or \emph{stacking}---parameter, $C$, 
remains constant along the chain, {\it i.e.}, independent of curvature and twisting.\par
As a result, using the scalings and parameter values presented below, we 
obtain the dimensionless Hamiltonian, 
\begin{eqnarray} 
H & = & \sum_{n=-N}^{n=N} \left\{ \frac{1}{2} \dot{u}_n^2  + 
  \frac{C}{2} \left(u_{n+1} - u_{n} \right)^2 \right. \nonumber \\
  & & \quad + \left( e^{-u_n} -1 \right)^2 +
  \frac{1}{2} \displaystyle \left. \sum_{m} \right.^{\prime} 
      J_{nm} u_n u_m \Big\},\label{eq:hamil}
\end{eqnarray}
where the prime indicates $m \neq n$ in the last summation,  accounting for the 
long-range interaction (LRI) between the dipoles. The total number of sites 
are thus $N_T = 2N+1$. Without the LRI, this Hamiltonian has previously been 
used to describe thermal denaturation in DNA, see Ref.~\cite{PRE 47/1}.\par
The LRI is a dipole-dipole interaction, with coefficients $J_{nm}$ given by 
\cite{PD 163, Landau Lifshitz}
\begin{equation} 
J_{nm} =  \frac{J 
 \left\{ {d}_n \cdot {d}_m - 
 	3 \left({d}_n \cdot {r}_{nm} \right) \left({d}_m 
 	\cdot {r}_{nm} \right) \right\}}
{|{r}_n - {r}_m |^3}, \label{eq:dipole}
\end{equation}
where ${r}_n$ and ${d}_n$ are the position vector and the unit dipole 
vector at the $n$th site, respectively, and ${r}_{nm}$ denotes the unit 
vector from the $n$th to the $m$th site
\begin{equation} \label{eq:unitvec}
{r}_{nm} = \frac{{r}_n - {r}_m}{|{r}_n - {r}_m |}.
\end{equation}
We note that the geometry of the chain only comes into play 
through the LRI's \footnote{This is also the case for a nonrigid chain 
\cite{EPL 59}.}.\par
Dimensionless variables have been introduced as
\begin{displaymath}
u_n = \alpha \tilde{u}_n, \quad  t = \tilde{t}/t_0, \quad  {r}_n = 
       \tilde{{r}}_n/l, \quad \textrm{ and  } H=\tilde{H}/D, 
\end{displaymath}
where original physical variables are indicated by tildes. $l$ is the constant 
intersite distance between neighboring base pairs, $\alpha$ and $D$ are 
parameters in the dimensional Morse potential, $D \left( e^{-\alpha \tilde{u}_n} - 1 
\right)^2$, and the time constant is given by $t_0 = \sqrt{M / D \alpha^2}$, 
where $M$ is the mass of a base pair.
The dimensionless parameters $C$ and $J$ are given by
\begin{equation} \label{eq:temp}
 C =  \tilde{C}/(D\alpha^2) \quad  \textrm{ and } \quad
 J = 2 \tilde{J}_{0} / l^3 D \alpha^2,
\end{equation}
with $\tilde{J}_0 = q^2/4 \pi \varepsilon_0$, where $q$ denotes the dipole 
charge of the base pair and $\varepsilon_0$ is the dielectric constant. \par 
\subsection{Parameter values \label{sec:param}}
We use the parameter values $D=0.04$ eV ($=0.64 \times 10^{-20}$ J), 
$\alpha=4.45 \textrm{ \AA}^{-1}$ ($=4.45 \times 10^{10}$ m$^{-1}$), 
$M=300 \textrm{ a.m.u.}$ ($=5.00 \times 10^{-25}$ kg) and the coupling parameter 
$\tilde{C}=0.06$ eV/$\textrm{\AA}^2$ ($=0.96$ J/m$^{2}$),
which have been widely used in DNA-like models
\cite{PRE 47/1, PLA 299, PD 163}.  The dimensionless 
stacking parameter then becomes $C=0.075$, which we use throughout the paper. 
We note that
$\tilde{C}$-values between $0.003$ eV/$\textrm{\AA}^2$ \cite{PRL 62} and 
$31.7$ eV/$\textrm{\AA}^2$ \cite{PD 126} have been reported in the literature.

The resulting time constant, $t_0 = 0.20$ ps, is in 
the picosecond range, as seen in Table~\ref{tab:param_2}.\par 
The dipole moment for the base pair, $d_{\textsf{dip}}$, is the geometric 
sum of the moments for the bases, which range between 3 and 7 debye 
\cite{JACS 113}. In our simulations we use $d_{\textsf{dip}} \approx 7$ debye.
The corresponding dipole charge $q = d_{\textsf{dip}} / a_{\textsf{dip}}$, 
where $a_{\textsf{dip}}$ is the equilibrium distance between the dipole charges 
($\approx 2$ \AA \cite{PD 163}) then becomes $1.17 \times 10^{-19}$ C, yielding 
$J=0.5$ \footnote{Recent calculations by J. Cuevas, J.F.R. Archilla, 
E.B. Starikov, and D. Hennig indicate a value of $q$ about 10 times smaller.}.
\par
\begin{table}
\begin{tabular}{ccc}
\hline
\hline
 Symbol & Parameter & Physical value \\
\hline
$\tilde{C}$ & stacking & $0.96$ J/m$^2$\\
$D$ & Morse depth & $0.64 \times 10^{-20}$ J\\
$\alpha$ & inverse Morse width & $4.45 \times 10^{10}$ m$^{-1}$\\
M & base pair mass & $5.00 \times 10^{-25}$ kg\\
$q$ & dipole charge & $1.17 \times 10^{-19}$ C\\
$\tilde{J}_{0}$ & interaction strength & $0.90 \times 10^{-28}$ Jm\\
$l$ & lattice constant & $3.4 \times 10^{-10}$ m\\
$t_0$ & time constant & $0.20 \times 10^{-12}$ s\\
\hline
\hline
\end{tabular} 
\caption{ \label{tab:param_2}Physical parameters for the DNA molecule.}
\end{table}
\subsection{Geometry \label{sec:geom}}
We introduce the twist angle, $\phi_n$, which the dipoles create with the 
$z$ direction, in the form of a kink (\ref{eq:twist}), where $\gamma$ gives 
the position of the kink center,
\begin{equation} \label{eq:twist}
\phi_n = 2 \arctan \left[ e^{- \tau \left( n-\gamma \right)} \right].
\end{equation}
Thus, $\gamma$ is 0 in the on-site case, but $\gamma = 1/2$ in the intersite 
case. 
The most important difference between the two is that the on-site case always 
has a dipole aligned in the bending plane at $n=0$ ($\phi_0 = \pi/2$).\par
\begin{figure}[h]
  \psfrag{x1}[cc][cc]{$-\!10$}
  \psfrag{x2}[cc][cc]{$-\!5$}
  \psfrag{x3}[cc][cc]{$0$}
  \psfrag{x4}[cc][cc]{$5$}
  \psfrag{x5}[cc][cc]{$10$}
  \psfrag{y1}[cc][cc]{$0$}
  \psfrag{y2}[cc][cc]{$\frac{\pi}{2}$}
  \psfrag{y3}[cc][cc]{$\pi$}
  \psfrag{Site}[tt][tc]{Shifted site, $n-\gamma$}
  \psfrag{Twist}[Bc][Bc]{Twist angle, $\phi_n$}
 \includegraphics[width=6cm]{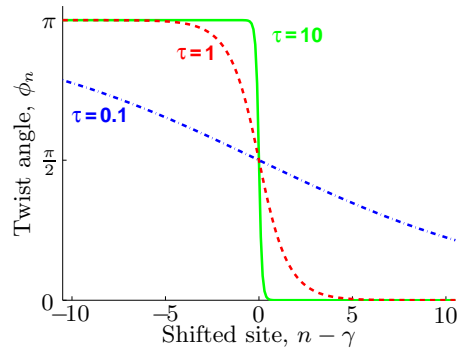}
 \caption{\label{fig:twist} (Color online) The twist angle, $\phi_n$, for 
 various values of the twist, $\tau$, around the chain center. On-site case, 
 $\gamma = 0$, intersite case, $\gamma = 1/2$.}
\end{figure}

As illustrated in Fig.~\ref{fig:twist}, the larger \emph{twist}, $\tau$, the 
faster the twist angle changes --- especially in the region of maximal twist 
and curvature. 
Note that the maximal slope of the twist function occurs at $\gamma$. 
In the limit $\tau \rightarrow \infty$, Eq.~(\ref{eq:twist}) gives 
$\phi_n = \pi$ for $n < 0$, and $\phi = 0$ for 
$n\geq 1$ for both cases. 
In the same limit, at $n=0$, the on-site case has 
$\phi_0 = \pi/2$, whereas the inter-site case has $\phi_0 = \pi$. In the limit 
$\tau \to 0$, all $\phi_n = \pi/2$ in both cases.\par
On the parabola embedded chain, 
$y_n = \left( \kappa/2 \right) \left( x_n-\gamma \right)^2$, the resulting 
unit dipole vectors then become
\begin{equation} \label{eq:d}
  {d}_{n} = \big( - \xi_n \kappa \left( x_n-\gamma \right) \sin \phi_n, 
  		\xi_n \sin \phi_n, 
  \cos \phi_n  \big),
\end{equation}
with $\xi_n = 1/\sqrt{1+\kappa^2 \left( x_n -\gamma \right)^2}$ (corresponding 
to the dark grey arrows in Fig.~\ref{fig:dnachain}). In the on-site case, 
$x_0 - \gamma \equiv 0$, and the other site positions are numerically 
calculated to fulfill the requirement that the distance between adjacent sites 
is always unity. In the intersite case, we define $x_0 - \gamma \equiv -1/2$ 
and $x_1 - \gamma \equiv 1/2$ and compute the other sites in a similar way. 
Thus, the axis of symmetry passes through the site $n=0$ in the on-site, 
$\gamma = 0$, case 
and passes through the middle of the bond connecting sites $n=0$  
and $n=1$ in the intersite, $\gamma = 1/2$, case
\footnote{In the intersite case, 
though, $x_N = x_{-N-1}$ and the legs are not complete symmetric in order to 
use the same initial conditions for both cases.}. See Fig.~\ref{fig:dnachain}.
\par
It is difficult to determine the actual value of the twist, $\tau$. The 
standard value for the twist angle between neighboring base pairs in an 
undisturbed DNA molecule is about $36^{\circ}$ \cite{PLA 253, JBP 24, PRE 66}. 
Requiring that a twist of $180^{\circ}$ should be achieved over five sites, 
thus corresponds to a value of $\tau$ about 1 (Fig.~\ref{fig:twist}) in 
equilibrium and larger when twisted more.\par
Similarly, the bending of a DNA chain can be approximately determined from 
experimental results. In Refs.~\cite{PLA 299, Cond 13} a parabolic 
approximation 
with curvature parameter $\kappa \leq 4$ was used. Here, we shall only 
consider the range  $\kappa \leq 2$ as larger curvature has never been observed 
in measurements.\par
\section{\label{sec:num}Simulation results}
From the Hamiltonian (\ref{eq:hamil}) we obtain the equations of motion 
\begin{eqnarray}
   \ddot{u}_n + C \left( 2u_n - u_{n-1} - u_{n+1} \right) & & \nonumber \\
  - 2 e^{-u_n} \left( e^{-u_n} - 1 \right) + 
   \left. \sum_{m} \right.^{\prime} J_{nm} u_m & = & 0. 
   \label{eq:moteqn}
\end{eqnarray}
\par
In the following we solve these equations numerically using a fourth order 
Runge-Kutta solver with free boundary conditions $u_{-N-1}=u_{-N}$ and 
$u_{N+1}=u_N$. In all 
simulations the relative change of the Hamiltonian is less than $10^{-5}$ and we
 consider a chain with $N_T=99$ sites.\par
In our previous work considering an approximate dipole-dipole interaction on a 
wedge shaped chain \cite{Funnel} we investigated random initial conditions. 
This was done as nonlinear excitations are known to be generated from 
randomness \cite{Muto, PD 57, PRE 47/1, PRE 60, PD 119}. As the outcome of 
collisions between nonlinear excitations depends strongly on their relative 
phases, we made 500 different random initial conditions to 
be able to find effects independent of the random phases.
The random initial conditions created nonlinear excitations, which led to 
bubble generation at various collision sites.
We found that bending of the chain caused the bubble generation to 
localize at the bent region.\par
In the following we use a similar approach to investigate the relationship 
between twist and curvature in this 
realistic expression for the dipole interaction (\ref{eq:dipole}). 
We consider random initial conditions: Initial displacements set to zero, 
{\it i.e.}, $u_n(0)=0$ for all $n$, while initial velocities of the chain 
sites are  normally distributed with mean value 
$\langle \dot{u}_n(0) \rangle =0$ and standard deviation $\sigma_{\dot{u}_n}$. 
The standard deviation is chosen to be 
$\sigma_{\dot{u}_n}=1.156$, corresponding a temperature of 
$T \approx 310$ K.\par 
The system dynamics is simulated for 100 different realizations of the initial 
conditions with stepsize $0.01$ in time. The simulations run for 100 time units 
(corresponding to $20$ ps) or until the Hamiltonian is no longer conserved. 
As mentioned, the constant dipole-dipole interaction coefficient $J=0.5$ is used in
all the simulations.\par
\begin{figure}[h] 
      \psfrag{y5}[cr][cr]{$80$}
      \psfrag{y4}[cr][cr]{$60$}
      \psfrag{y3}[cr][cr]{$40$}
      \psfrag{y2}[cr][cr]{$20$}
      \psfrag{y1}[cr][cr]{$0$}
      \psfrag{x1}[tc][cc]{$0$}
      \psfrag{x2}[tc][cc]{$5$}
      \psfrag{x3}[tc][cc]{$10$}
      \psfrag{x4}[tc][cc]{$15$}
      \psfrag{Displacement}[cc][tc]{Displacement [{\AA}]}
      \psfrag{Time}[cc][bc]{Time [ps]}
    \centerline{\epsfig{file=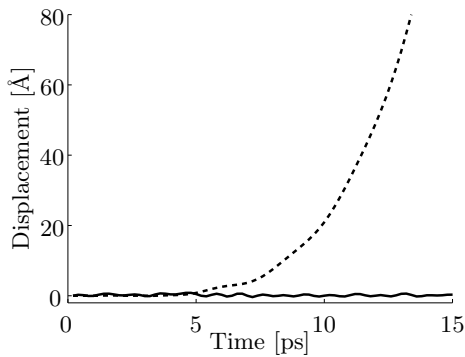, width=60 mm}}
  \caption{Evolution of the center site amplitude for $\kappa=1.0$ and 
  	$\tau=6$ (dashed curve) and $\tau=4$ (solid curve) for an on-site 
	case.}
  \label{fig:BU}
\end{figure}

%
%
For different values of twist or curvature, the same initial condition can 
result in very different behavior. Consider Fig.~\ref{fig:BU}, where the same 
initial condition is simulated for two different values of the twist, 
$\tau$, in the on-site case. With the larger twist (dashed curve) the 
amplitude grows in an exponential-like manner (see Sec.~\ref{sec:anal}).
\par
At the end of each simulation, we examine the amplitudes. We use a threshold 
value of $u_n=100$, corresponding to about $20$ {\AA}, {\it i.e.}, twice the 
equilibrium distance between base pairs. If the threshold is exceeded in at 
least two adjacent sites, we consider this as a precursor for bubble generation.
In the following sections, we have depicted the regions of 
curvature, $\kappa$, and twist, $\tau$, where at least one of the 100 
simulations result in bubble generation.\par
\subsection{\label{sec:onsite}Results for the on-site case}
\begin{figure}[h] 
      \psfrag{y4}[cr][cr]{$10$}
      \psfrag{y3}[cr][cr]{$5$}
      \psfrag{y2}[cr][cr]{$3$}
      \psfrag{y1}[cr][cr]{$2$}
      \psfrag{x1}[tc][cc]{$0$}
      \psfrag{x2}[tc][cc]{$0.5$}
      \psfrag{x3}[tc][cc]{$1$}
      \psfrag{x4}[tc][cc]{$1.5$}
      \psfrag{x5}[tc][cc]{$2$}
      \psfrag{(aa)}[cr][cr]{(a)\hspace{6mm}}
      \psfrag{(bb)}[cc][cc]{(b)}
      \psfrag{(c)}[cc][cc]{(c)}
      \psfrag{tau}[Bc][tc]{Twist, $\tau$}
      \psfrag{kappa}[tc][bc]{Curvature, $\kappa$}
  \centerline{\epsfig{file=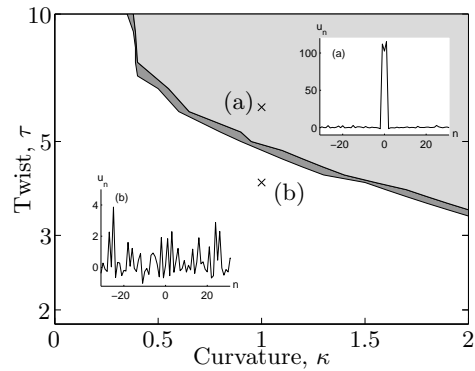, width=6 cm, angle=0}}
  \caption{Region of bubble generation in the on-site case. Light grey region, 
  	   bubble generation. 
	   Dark grey region, transition region indicating the uncertainty of the 
	   simulations. White region, no bubble generation. Insets show 
	   the displacement, $u_n$, versus site, $n$, at the points (a) and (b) at
	   simulation times (a) $t=51$ and (b) $t=100$ for the same initial 
	   condition.}
  \label{fig:Lkappa_onsite}
\end{figure}

The combined effect of both curvature and twist for the on-site case is depicted 
in Fig.~\ref{fig:Lkappa_onsite}, where the shaded region corresponds to bubble 
generation. We see that bubbles are 
generated for strong twist and strong curvature, which is expected here. For 
the on-site case, the strongest attraction between dipoles occur at the 
sites $n=-1$ and $n=1$ (which are almost antiparallel for strong twist). 
Increasing the curvature bring these next-to-center sites closer, which 
augments the dipole-dipole interaction (\ref{eq:dipole}). As the inset (a) 
shows,
the amplitude increase is localized to the region of maximal twist and curvature. 
We are aware that in reality, whole regions of DNA base pairs move apart during 
denaturation. Therefore, our mechanism should only be perceived as a precursor 
for bubble generation.\par
The exact shape of Fig.~\ref{fig:Lkappa_onsite} depends on the chosen 
amplitude threshold---as well as system parameters---but its qualitative 
shape is unchanged. Close to the region 
border, only a few of the simultations results in bubble generation, but this 
number increases as one proceeds in the direction of stronger twist and larger 
curvature ({\it i.e.}, towards the upper right corner). We note that for the 
on-site 
case the amplitude is increased at the three center sites $n=-1$, $n=0$, and $n=1$ 
as indicated in the inset (a). 

\par
\subsection{\label{sec:intersite}Results for the intersite case}
\begin{figure}[h] 
      \psfrag{y7}[cr][cr]{$2.5$}
      \psfrag{y6}[cr][cr]{}
      \psfrag{y5}[cr][cr]{}
      \psfrag{y4}[cr][cr]{}
      \psfrag{y3}[cr][cr]{}
      \psfrag{y2}[cr][cr]{$2.0$}
      \psfrag{y1}[cr][cr]{}
      \psfrag{x1}[tc][cc]{$0$}
      \psfrag{x2}[tc][cc]{$0.5$}
      \psfrag{x3}[tc][cc]{$1$}
      \psfrag{x4}[tc][cc]{$1.5$}
      \psfrag{x5}[tc][cc]{$2$}
      \psfrag{(cc)}[cc][cc]{(a)}
      \psfrag{(dd)}[cr][cr]{(b)\hspace{6mm}}
      \psfrag{tau}[Bc][tc]{Twist, $\tau$}
      \psfrag{kappa}[tc][bc]{Curvature, $\kappa$}
  \centerline{\epsfig{file=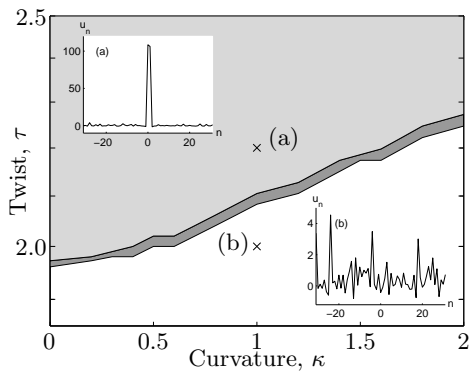, width=6 cm, angle=0}}
  \caption{Region of bubble generation in the intersite case. Light grey region, 
  	   bubble generation. 
	   Dark grey region, transition region indicating the uncertainty of the 
	   simulations. White region, no bubble generation. Insets show 
	   the displacement, $u_n$, versus site, $n$, at the points (a) and (b) at 
	   simulation time (a) $t=45$ and (b) $t=100$ for the same initial 
	   condition.}
  \label{fig:Lkappa_intersite}
\end{figure}

In the intersite case, Fig.~\ref{fig:Lkappa_intersite}, the picture is 
different. First of all, the twist needed for 
bubble generation is smaller. Second, the dependence 
on the curvature is less pronounced. This is because the dipoles, that for 
strong twist are antiparallel, in this case are neighbors. Since, in the 
framework of our model, the chain has 
constant distance between adjacent sites, increasing the curvature does not 
increase the tendency for bubbles to be generated. \par
  In fact, the opposite is the case: As the dipole twist is perpendicular to 
the chain, increasing the curvature has the consequence that the center dipoles 
interact in a less attracting way, since the center dipoles 
become more antiparallel (Fig.~\ref{fig:SiteEn_taus}).\par
 Note that 
in this case, the displacement at the two center cites, $n=0$ and $n=1$, is 
increased as the inset (a) shows. Close to the region border, the number of 
simulations resulting in bubble generation is small, but it increases as one 
increases the twist or decreases the curvature ({\it i.e.}, moves to the upper 
left corner).%
\par
\subsection{Effective potential}
\begin{figure}[h] 
  \centerline{
      \psfrag{y3}[cr][cr]{\scriptsize $0.0$}
      \psfrag{y2}[cc][cc]{\scriptsize }
      \psfrag{y1}[cr][cr]{\scriptsize $-1.0$}
      \psfrag{x1}[tc][cc]{\scriptsize $0$}
      \psfrag{x2}[tc][cc]{}
      \psfrag{x3}[tc][cc]{}
      \psfrag{x4}[tc][cc]{}
      \psfrag{x5}[tc][cc]{}
      \psfrag{x6}[tc][cc]{\scriptsize $10$}
      \psfrag{(a)}[cc][cc]{(a)}
      \psfrag{(b)}[cc][cc]{(b)}
      \psfrag{Energy}[Bc][tc]{$\ V_n - V_{\infty}$}
      \psfrag{Twist}[tc][bc]{Twist, $\tau$}
    \epsfig{file=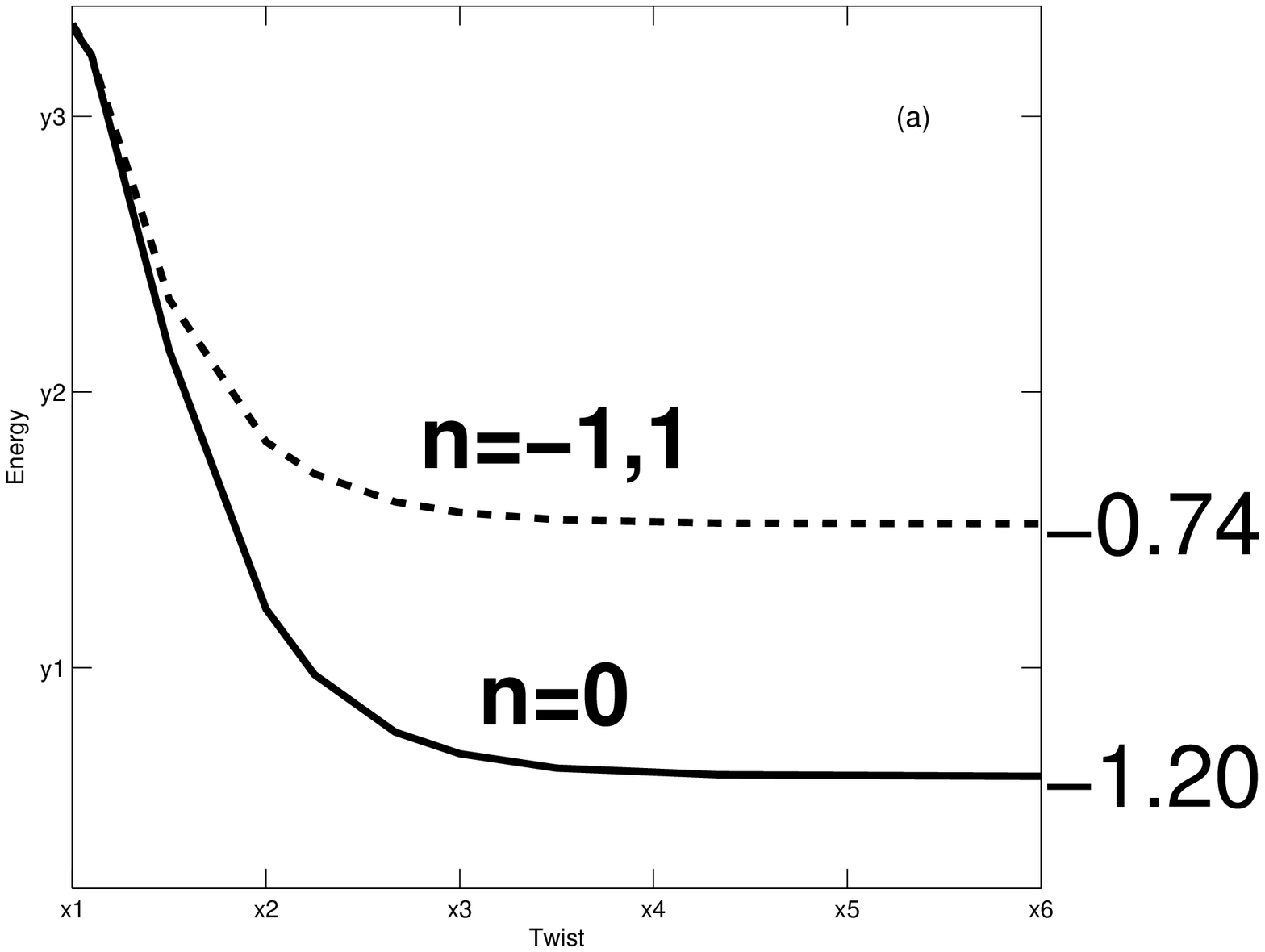, width=38 mm, height=30mm, angle=0}
    \hspace{4mm}
    \epsfig{file=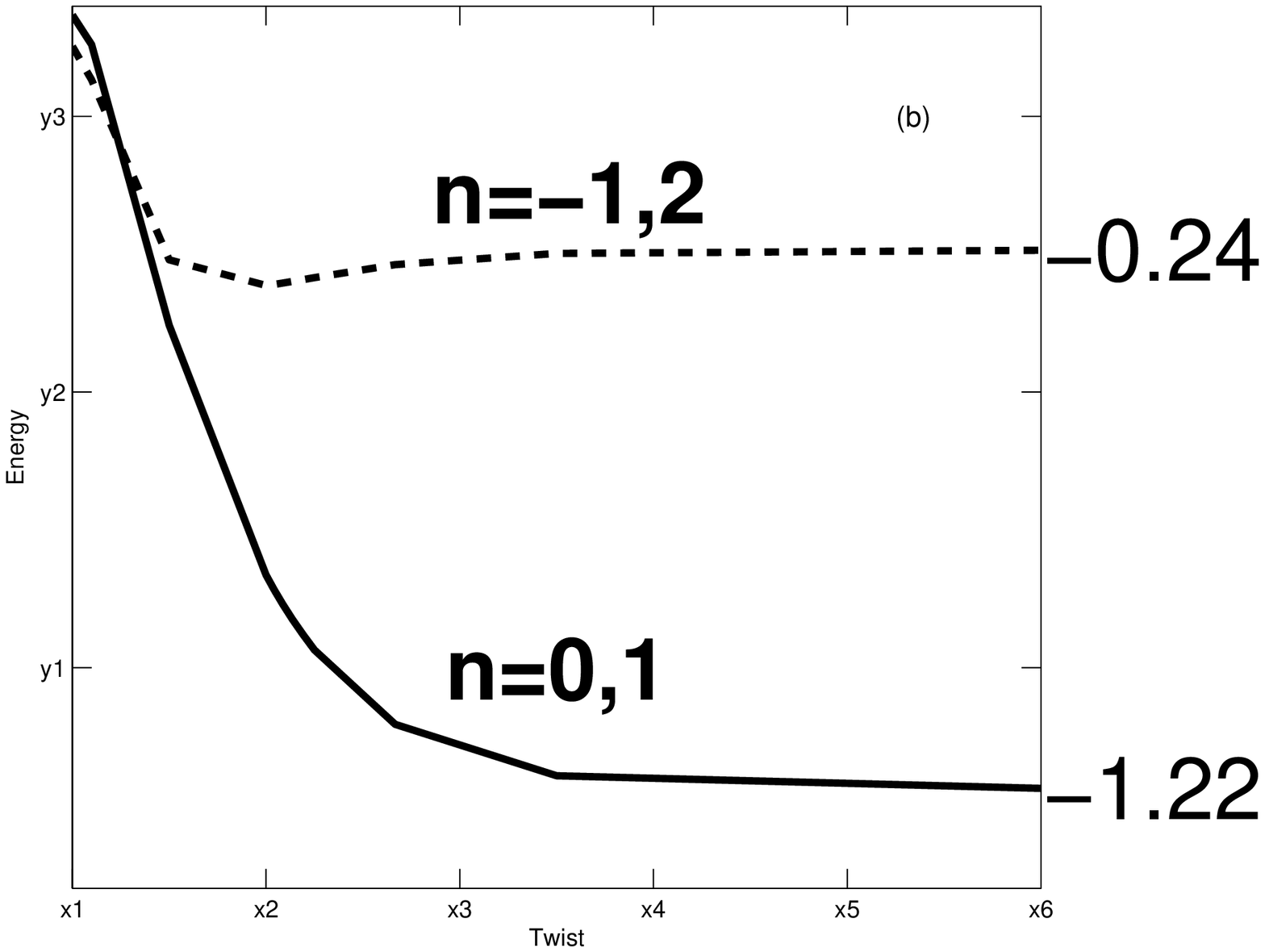, width=38 mm, height=30mm, angle=0}}
  \caption{Effective potential 
  	$V_n - V_{\infty}$ for 
  	   constant curvature $\kappa = 1.0$. 
           (a) On-site case. (b) Intersite case.}
  \label{fig:SiteEn_kap10}
\end{figure}

It is obvious that having initially randomly distributed energy along the chain, 
successful bubble generation should include, as a first stage, funneling of 
energy in the bent and twisted region. Therefore we can expect that only in the 
case when this region acts as a potential well, bubbling may occur. 
The behavior found in 
Figs.~\ref{fig:Lkappa_onsite} and \ref{fig:Lkappa_intersite} can be 
qualitatively explained by the \emph{effective on-site potential}, 
$V_{n} \equiv \sum_{m}^{\prime} J_{nm}$, which is introduced as
\begin{eqnarray*}
  \sum_n \left. \sum_{m} \right.^{\prime} J_{nm} u_n u_m & = 
  & -\frac{1}{2} \sum_n \left. \sum_{m} 
     \right.^{\prime} J_{nm} \left( u_n - u_m \right)^2 \\ 
  & & \quad + \sum_n V_{n} u_n^{2}.
\end{eqnarray*}
We consider the dipole potential at given sites for both the on-site and the 
intersite case for constant curvature, $\kappa=1.0$, with respect to 
the ``ground state'' at $n \to \pm \infty$. 
Thus, Fig.~\ref{fig:SiteEn_kap10} 
depicts the depth of the potential well for constant curvature for both the 
on-site [Fig.~\ref{fig:SiteEn_kap10}(a)] and the intersite 
[Fig.~\ref{fig:SiteEn_kap10}(b)] case .
We see that in the vicinity of the bending point, there exist an effective 
potential well for $\tau$ larger than about $0.5$. This corresponds to the 
behavior 
found in Figs.~\ref{fig:Lkappa_onsite} and \ref{fig:Lkappa_intersite}.
The depth of the potential well increases 
with increasing twist, until a saturation is reached at $\tau \approx 5$. 
The existence of a potential well is a necessary, but not sufficient, 
condition for bubble generation.
\par
\begin{figure}[h] 
  \centerline{
      \psfrag{y3}[cr][cr]{\scriptsize $0.0$}
      \psfrag{y2}[cc][cc]{\scriptsize }
      \psfrag{y1}[cr][cr]{\scriptsize $-1.0$}
      \psfrag{x1}[tc][cc]{\scriptsize $0$}
      \psfrag{x2}[tc][cc]{}
      \psfrag{x3}[tc][cc]{}
      \psfrag{x4}[tc][cc]{}
      \psfrag{x5}[tc][cc]{\scriptsize $2$}
      \psfrag{(c)}[cl][cc]{(a)}
      \psfrag{(d)}[cl][cc]{(b)}
      \psfrag{Energy}[Bc][tc]{$\  V_n - V_{\infty}$}
      \psfrag{Curvature}[tc][br]{Curvature, $\kappa$}
    \epsfig{file=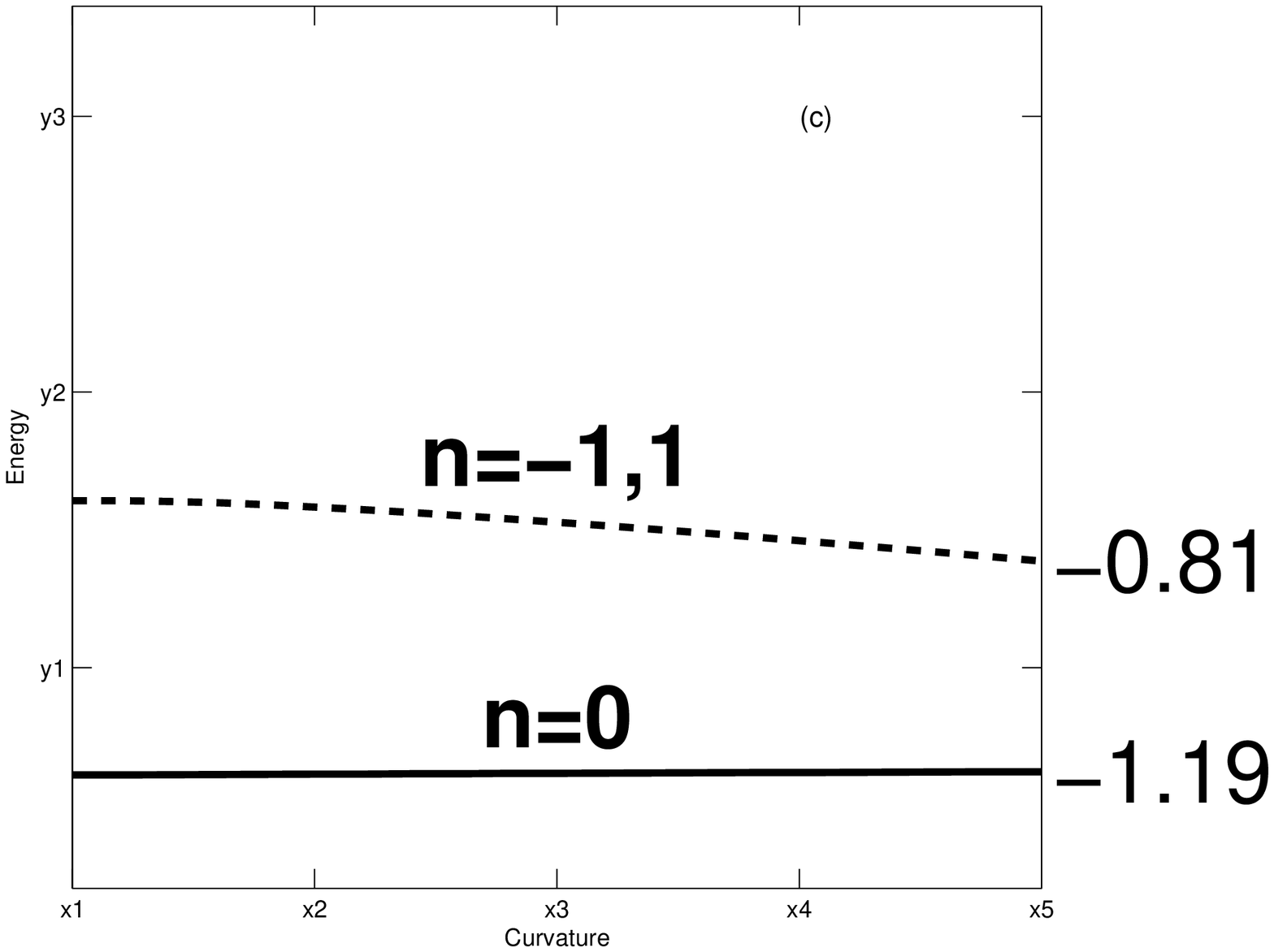, width=38 mm, height=30mm, angle=0}
    \hspace{4mm}
    \epsfig{file=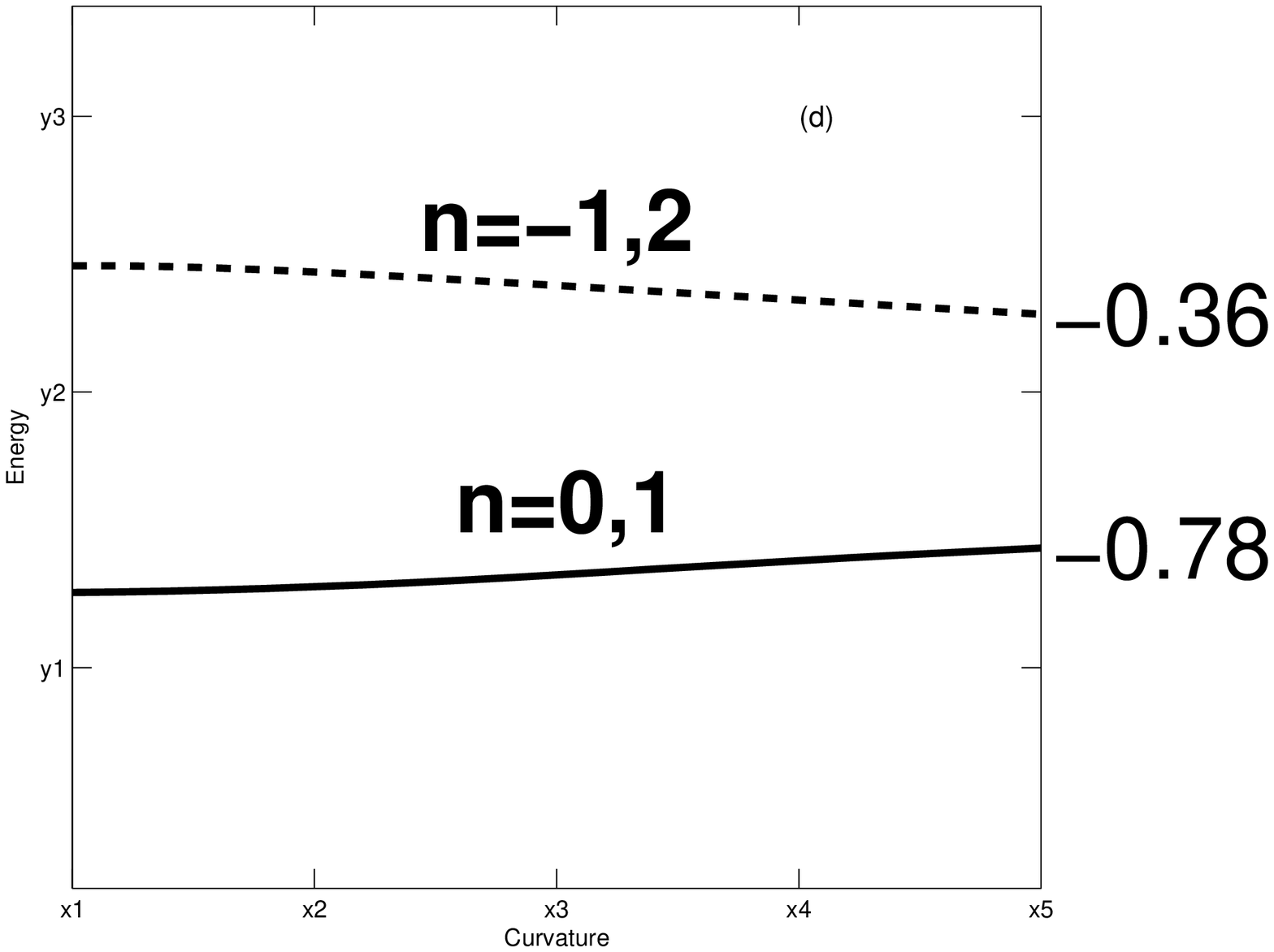, width=38 mm, height=30mm, angle=0}}
  \caption{Effective potential $V_n - V_{\infty}$ for constant twist. 
           (a) On-site case, $\tau = 6$.
	   (b) Intersite case, $\tau = 2$.}
  \label{fig:SiteEn_taus}
\end{figure}

For constant twist, Fig.~\ref{fig:SiteEn_taus}, we find different behaviors 
in the two chain configurations. For the on-site case with $\tau = 6$ 
[Fig.~\ref{fig:SiteEn_taus}(a)], 
we see a decreasing potential well depth with increasing curvature at sites 
$n=-1$ 
and 
$n=1$, whereas the potential at $n=0$ is almost constant. This corresponds to 
an effective potential well for increasing curvature in the on-site case. 
In the intersite case [Fig.~\ref{fig:SiteEn_taus}(b)], the twist is fixed at 
$\tau = 2$, but 
in contrast to the on-site case, we see that the depth of the potential well
increases with increasing curvature at the center sites $n=0$ and $n=1$. 
Therefore, bubble generation is not found for larger 
curvature in the intersite case, corresponding to the behavior seen in 
Fig.~\ref{fig:Lkappa_intersite}.
\par
Thus both curvature and twisting play a role in the localized formation of 
precursors for
denaturation bubbles in our model of the DNA molecule and it is clear that 
the chain configuration is important. Stronger twist increases the initiation 
of bubble generation in both cases considered. The effect of increasing 
curvature is different: In the on-site case, curvature clearly 
enhances bubble generation, but in the intersite case it slightly decreases 
the formation of bubbles in the range of $\kappa$ considered.\par
Simulations for smaller values of the parameter $J$ showed that stronger twist 
was required to create bubbles.\par
\section{Analytical approach\label{sec:anal}}
Considering only the center sites $n=0$ and $n=1$ in the intersite case, we 
are in effect looking at a \emph{dimer}. Assuming that both displacements are 
equal, $u_0 = u_1 \equiv u$, the coupling term in the Hamiltonian, 
Eq.~(\ref{eq:hamil}), vanishes and we are left with
\begin{equation} \label{eq:Hanal}
H = \dot{u}^2 + 2 \left( e^{-u} -1 \right)^2 - J u^2,
\end{equation}
\noindent with $J = | J_{01} | = | J_{10} |$. For a strong twist, $\tau \gg 1$, 
the last term becomes negative, due to opposite dipole orientations, 
corresponding to attractive interaction. The effective potential 
\begin{equation} \label{eq:veff}
V(u) = 2 \left( e^{-u} -1 \right)^2 - J u^2  
\end{equation}
is shown as the dashed curve in Fig.~\ref{fig:Morse}.\par 
Equation~(\ref{eq:Hanal}) may now be integrated as
\begin{equation} \label{eq:init}
t-\bar{t} = \int_{\bar{u}}^u \frac{d w}{\sqrt{H -  
2 \left( e^{-w} - 1 \right)^2 + J w^2 }},
\end{equation}
\noindent where $u=\bar{u}$ at the time $t=\bar{t}$. Choosing $\bar{t}$ so 
large that $w \gg 1$, the power term in the square root of 
Eq.~(\ref{eq:init}), $Jw^2$, dominates (as seen in Fig.~\ref{fig:Morse}). 
Therefore, Eq.~(\ref{eq:init}) may be approximated as 
\begin{displaymath}
t-\bar{t} \approx \int_{\bar{u}}^u \frac{dw}{\sqrt{J w^2}},
\end{displaymath}
\noindent from which we find $u \varpropto \exp 
\left[ \sqrt{J}(t-\bar{t}) \right]$, in accordance with the exponential 
behavior found numerically in Fig.~\ref{fig:BU}. 
A similar approach can be used for the on-site case with identical results.
\par
\section{Conclusion\label{sec:concl}}
We have shown that bubble generation in DNA-like models can be initiated by 
curvature and twisting of the molecular strands.\par
Stronger twist facilitate bubble generation, whereas the effect of curvature
depends on the details of the geometry. 
For the 
on-site case, increasing curvature increases the 
tendency for bubble generation. 
Conversely, the intersite case decreases bubble generation for increasing 
curvature. \par
Bubbles emerge in the region of maximal twist and curvature, and are found at 
physiological temperatures.\par
Numerical results are supported by analytical approximation. 
Widely used parameter values for DNA are used in our model.\par
\begin{acknowledgments}
The authors wish to thank P. Fischer and N.C. Albertsen, Informatics and 
Mathematical Modelling, Technical University of Denmark, for helpful and 
inspiring contributions. Lars Hemmingsen, Department of Physics, 
Technical University of Denmark is acknowledged for 
Ref.~\cite{JACS 113}. One of the authors (Yu.B.G.) thanks Informatics and Mathematical 
Modelling for hospitality and a Guest Professorship as well as support 
from the research center of quantum medicine ``Vidguk''. 
The third author (O.B.) acknowledges support 
from the Danish Technical Research Council (Grant No.~26-00-0355). The work 
is supported by LOCNET Project No.~HPRN-CT-1999-00163. 
\end{acknowledgments}
\end{document}